*CORRESPONDENCE
Quentin A. Parker,
quentinp@hku.hku

SPECIALTY SECTION
This article was submitted to Low-Temperature Plasma Physics, a section of the journal Frontiers in Astronomy and Space Sciences

RECEIVED 13 March 2022
ACCEPTED 15 June 2022
PUBLISHED 29 June 2022

CITATION
Parker QA (2022), Planetary nebulae and how to find them: A concise review. *Front. Astron. Space Sci.* 9:895287. doi: 10.3389/fspas.2022.895287

# Planetary nebulae and how to find them: A concise review

Quentin A. Parker [1,2]*

[1]Laboratory for Space Research, University of Hong Kong, Hong Kong, Hong Kong SAR, China,
[2]Department of Physics, The University of Hong Kong, Hong Kong, Hong Kong SAR, China

This review provides useful background and information on how we find, vet and compile Planetary Nebulae (PNe) candidates and verify them. It presents a summary of the known Galactic PNe population and their curation in the Hong Kong/AAO/Strasbourg/H$\alpha$ PNe catalogue, "HASH". It is a simple introduction for anyone interested in working with PNe, including postgraduate students entering the field and for more general interest too.

KEYWORDS

**Planetary nebulae, wide-field surveys, narrow-band imaging, catalogues, databases, spectroscopy, diagnostics, Planetary nebula**

## 1 Introduction

A key question in any review into the Planetary nebulae (PNe) phenomenon is why do we care about studying them? What is their astrophysical value? Put simply, they are open windows into late stage stellar evolution. PNe are nothing to do with planets, only resembling gas giants in our own Solar System in early telescopes. They are the ejected, ionised envelopes of low-to intermediate-mass, dying stars and they are some of the most beautiful and complex of astrophysical sources. They are also amongst the most important to understand. This is because, despite their short lifetimes, typically ~25,000 ± 5,000 years (Jacob et al., 2013; Badenes et al., 2015) they provide unique insights into late-stage evolution of stars in the mass range $\sim 1 - 8\ M_{\odot}$. They are important probes of nucleosynthesis processes, mass-loss physics and post AGB timescales (Iben, 1995; Miller Bertolami, 2016), and Galactic enrichment and abundance gradients (Maciel and Costa, 2003; Henry et al., 2010; Maciel et al., 2015; Stanghellini and Haywood, 2018). Their progenitor stars dominate all stars above one solar mass in our Galaxy. Hence PNe, via their ejecta, are responsible for a large fraction of mass return to and chemical enrichment of the interstellar medium. There has been confusion in the literature about how to actually define what constitutes the PN phenomena but Frew and Parker (2010) provide the solid working definition used here. Fundamentally, the gaseous shell is generally optically thin and ionised by the post-AGB star (or stars in the case of common envelope) from which the gas originally emanated. This is unlike for symbiotic systems, which are often confused with PNe, where the ionised gas comes from a binary companion and not the ionising star which may even have had its own true PN earlier.

In Figure 1 a now iconic collage, shows 22 well-known PNe artistically arranged in a spiral pattern by order of approximate physical size as calculated from the SB-r statistical distance scale of Frew et al. (2016) - see caption for further details.

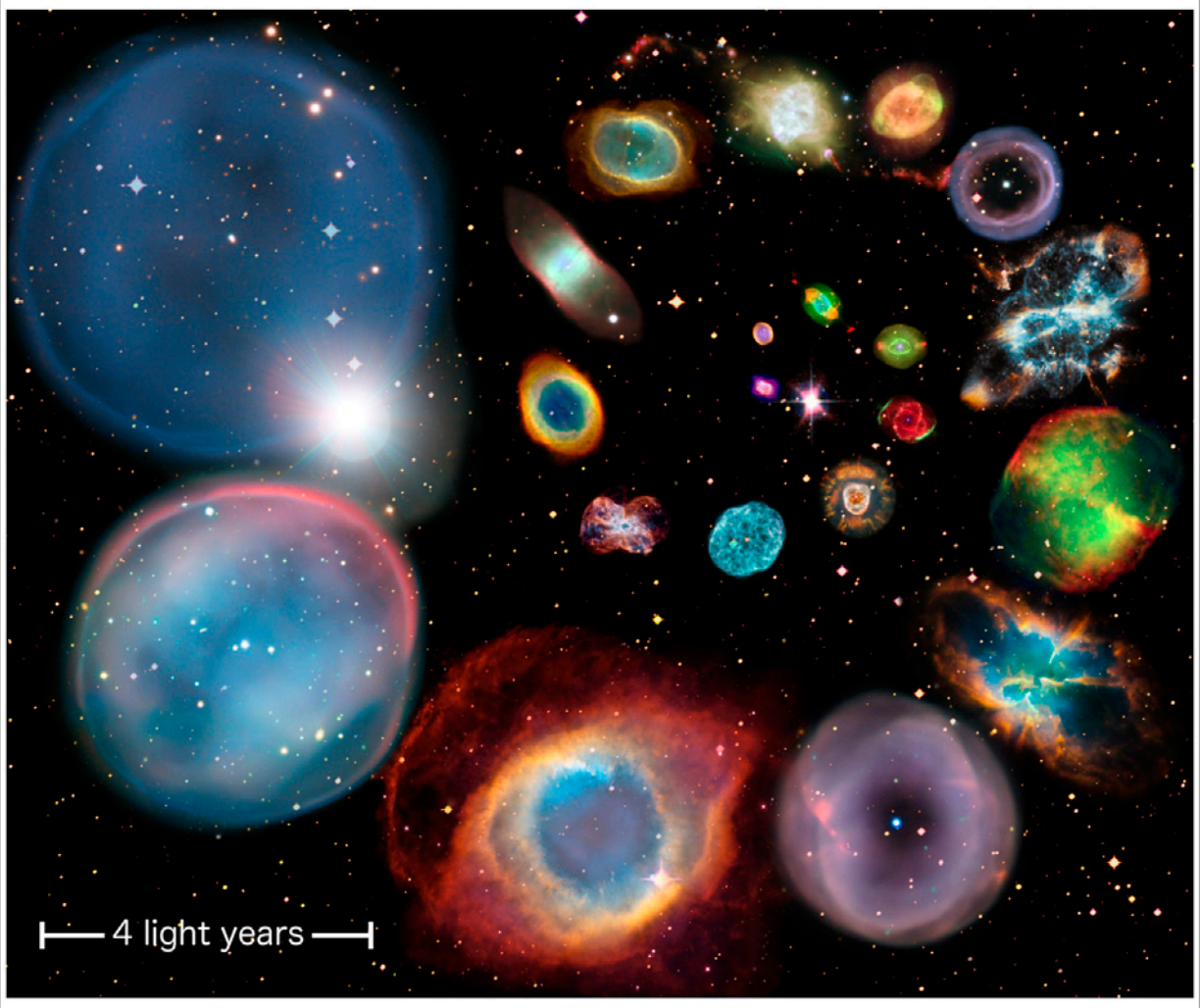

FIGURE 1
A now iconic collage showing 22 individual well-known PNe, artistically arranged in a spiral pattern by order of approximate physical size. Each nebula's size is calculated from the SB-r statistical distance scale from Frew et al. (2016). It can be applied to PNe exhibiting the entire range of surface brightness, morphology and size. The largest PNe have a surface brightness about a hundred thousand times fainter than the smallest and can reach up to 3 pc across. General image credit: ESA/Hubble and NASA, ESO, NOAO/AURA/NSF from an idea by the author and Ivan Bojičić and rendered by Ivan Bojičić with input from David Frew and the author. The names of all these iconic PNe in the figure starting at the top left hand corner and following the spiral are: Abell 33, K 1–22, NGC 7293, IC 5148/50, NGC 2818, NGC 6853, NGC 5189, IC 4406, Shapley 1, IC 289, Fleming 1, NGC 3132, IC 4406, NGC 6720, NGC 2440, NGC 1501, NGC 2392, NGC 6543, NGC 6826, NGC 7009, IC 418, NGC 7027, HD 44179.

Of course there have been many books and conference proceedings, such as the IAU PNe Symposia series and Asymmetrical PNe meetings, dedicated to PNe. This is supplemented by large numbers of journal papers and articles concerning the PN phenomena, with over 8,000 published since 1900 and ~2200 since 1990. Together these encaspulate the combined wisdom of these fascinating objects. Selected specific PNe reviews include Miller (1974) and Balick and Frank (2002), while a review of all PNe research just from 2014 is given by Zijlstra (2015) while recent work on finding PNe in broad band surveys was reported by Vejar et al. (2019). Finally, I recommend the excellent, recent review of Kwitter and Henry (2022). This paper is not intended to compete with these materials for PNe but adopts a more practical approach to reviewing the techniques and processes used and available to actually find PNe and to use them for science with the help of the Hong Kong/AAO/Strasbourg/H$\alpha$ PNe catalogue (HASH) PNe database, Parker et al. (2016) and see Section 12 later.

## 2 Basic PNe properties

PNe exhibit specific physical and observational characteristics from the UV to radio regimes. Furthermore, their rich, strong, optical emission-line spectra are powerful laboratories for plasma physics where most energy comes from the lines. There is little continuum unless any central star of the PN (CSPN) is contained within the spectrograph slit/fibre. Such strong emission lines facilitate PNe detection to large distances, including the Magellanic clouds (Reid and Parker, 2006, 2013) and more remote, external galaxies (Magrini et al., 2016; Bhattacharya et al., 2019). Here, the bright end exponential cut-off to the ensemble brightness distribution (the "luminosity function" or PNLF) has been shown to be a powerful cosmological distance calibrator or "standard candle", capable of determining the Universe's scale to 10%, e.g., Jacoby (1997), Ciardullo et al. (2005; 2012), Gesicki et al. (2018), Kreckel et al. (2017) and Scheuermann et al. (2022).

PNe are, in fact, the main observable tracer of low-mass stars at larger distances. The emission lines also allow the determination and analysis of chemical abundances and permit the estimation of shell expansion velocities (Gesicki and Zijlstra, 2000) and ages, so probing the physics and timescales of stellar mass loss (Iben, 1995). The measured radial velocities can trace the kinematic properties of observed PNe, enabling us to determine if they belong to a younger or older stellar population, in, say, the Galactic Bulge (Beaulieu et al., 2000; Smith et al., 2017). Their kinematic properties and visibility also make PNe useful kinematical probes for understanding the structure of galaxies, and to test whether a galaxy contains a substantial amount of dark matter (Romanowsky et al., 2003). The PNe birth rate, e.g., Cahn and Wyatt (1976) and Peimbert (1993), also effectively gives the death rate of stars born billions of years ago. Their complex morphologies provide clues to their formation, evolution, mass-loss processes, ISM interactions (Sabin et al., 2012) the possible shaping role by magnetic fields (Leal Ferreira, 2014; Sabin et al., 2015), binary central stars (De Marco, 2009; Miszalski et al., 2009; Hillwig et al., 2016; Jones and Boffin, 2017), common envelopes (García-Segura et al., 2018) and even massive planets (Soker, 2006; Sabach and Soker, 2018; Hegazi et al., 2020). As the central star fades to become a white dwarf and the nebula expands, the integrated flux, surface brightness and radius change in ways that can be predicted by current stellar and hydrodynamic theory (Dopita and Meatheringham, 1991). Eventually PNe dissipate into the ISM but can reach surprisingly large physical sizes (up to 3 pc) while still remaining detectable as a coherent entity before they do, e.g., PFP 1 Pierce et al. (2004).

## 3 How many PNe are there in the galaxy?

In all the above ways PNe are powerful astrophysical tools. This makes them valuable targets for discovery in our own Galaxy,

the Local Group, and beyond. Indeed, although the current number of Galactic PNe known is around 3,800 (Parker et al., 2016), more than double what it was 15 years ago, this is still far short of our best estimates of the expected Galactic PNe population that ranges from ~6,600 to ~45,000 (Moe and De Marco, 2006). Such estimates are based on whether the binary PNe formation scenario is adopted (De Marco, 2009) giving the lower number, or if single stars alone can form them (my belief). Thereafter these estimates then depend on the population synthesis models or extrapolations from local space densities, e.g., see Pottasch, (1996) and Jacoby et al. (2010). This number has a direct bearing on our understanding of stellar evolution theory, Galactic chemical enrichment rates and Galactic ecology. However, we still need to improve the detection completeness for Galactic PNe to cover their full evolutionary and phenomenological diversity and how this relates to their stellar progenitor populations and properties. This, and their use as key Galactic probes, provides strong motivation to hunt for them. Much progress has been made since the major Galactic PNe discoveries report by Parker et al. (2006) and Miszalski et al. (2008), as well as obtaining a better understanding of the different evolutionary pathways that can lead to PNe formation, e.g., Stanghellini and Pasquali, (1995) and Frew and Parker (2012).

## 4 Brief background on early PNe catalogues

The first known observation of a PN, the famous "Dumbbell" nebula, M 27, was by Charles Messier in 1764. By 1800, 33 additional PNe had been added, primarily by William Herschel, though their true nature was not yet understood. Thereafter, the class was incrementally added to over the next two centuries as their nature came to be seen as the ionized ejecta from lower mass, evolved stars off the AGB en-route to the White Dwarf phase.

The first major PNe compilation was the catalogue of Perek and Kohoutek (1967), followed by the ESO PNe catalogues of Acker et al. (1992; 1996) and the equivalent catalogue of Kohoutek (2001). These catalogues represent heterogeneous samples of essentially optical PNe discoveries compiled from all sources, telescopes and spectrographs over the previous 200 years. They comprised between 1,000 and ~1,900 putative "True" and candidate PNe. These important, historical, long term cataloguing works of Galactic PNe have now been superseded and this forms the basis for much of this paper.

## 5 Correct PNe identification and the problem of mimics

The astrophysical utility of PNe is predicated on their correct identification. As an example Cohen et al. (2011) have shown that by using mid-infrared (MIR) and radio diagnostics, 45% of the putative "known" pre-MASH PNe in the GLIMPSE-I region (Churchwell et al., 2009) are likely contaminants, mostly compact HII regions.

Apart from multi-wavelength imagery and canonical PNe shapes, optical spectra remain an essential tool to complete preliminary identification where typical PNe spectral signatures are used. These include the collisionally excited "forbidden" emission lines of elements like Oxygen, Nitrogen, Neon, Argon and Sulphur and the standard recombination lines of Helium and Hydrogen.

In Figure 2 are presented four example spectra of bona-fide PNe that exhibit different spectral characteristics that represent the emission line and emission line ratio diversity seen in PNe. The spectra range from that of an evolved bipolar (top left) that suffers from extinction, to a very low excitation (VLE), compact, young PN (top right) with no detectable [OIII]. Optical HST imagery of some of these very compact VLE PNe reveal their PNe morphologies and often their CSPN, thus confirming their nature. A high excitation PN is shown at bottom left while finally, at bottom right, is the well known elliptical PN NGC 2022 (G196.6-10.9) where HeII 4686Å is as strong as H$\beta$. In the absence of obscuring dust, the [OIII] is often the strongest line in a PN's optical spectrum and is why so many early PNe discoveries were on B-band photographic plates. See figure caption for further details.

A key complicating point is that bona-fide PNe spectra, while having common characteristics, can also, as seen in Figure 2, exhibit significant differences. This is due to varying levels of excitation from the central ionising star at the observed stage of PN evolution. Spectra can range from so-called young, compact, VLE, dense-envelope PNe, where [OIII] is absent, the [SII] lines indicate high electron density and the [NII] to H$\alpha$ ratio typically $> 0.5$ (too high for any HII region); to high excitation PNe with HeII 4686Å emission in the blue, but only H$\alpha$ obvious in the red. Then there are the so called Type I PNe (Kingsburgh and Barlow, 1994) that often exhibit high ratios of [NII] to H$\alpha$, are usually bipolar and are thought to emerge from higher mass progenitors (so a younger population), e.g., Calvet and Peimbert, (1983) with higher nitrogen and helium abundances (Peimbert et al., 1995). They are unsurprisingly, therefore, typically found closer to the Galactic mid-plane. There is also a largely unknown population of dust obscured PNe where only faintly visible H$\alpha$ and [NII] emission lines are perhaps detectable in the far optical, or at even longer wavelengths in the near-infrared (NIR) where the [SIII] 9069 and 9532Å lines are strong. See Jacoby and Steene (2004) for work on hunting for obscured PNe in the centre of the Galaxy using a narrow-band [SIII] filter in the NIR near 9532Å to locate candidates for follow-up spectroscopy. Other dust obscured PNe have been found (Fragkou et al., 2018). This is after having been shown where to look for possible very faint detections in H$\alpha$ survey imagery using diagnostic radio data, from, for example, the Cornish Very Large Array (VLA) survey of the inner Galactic Plane at 5 GHz, see Hoare et al. (2012).

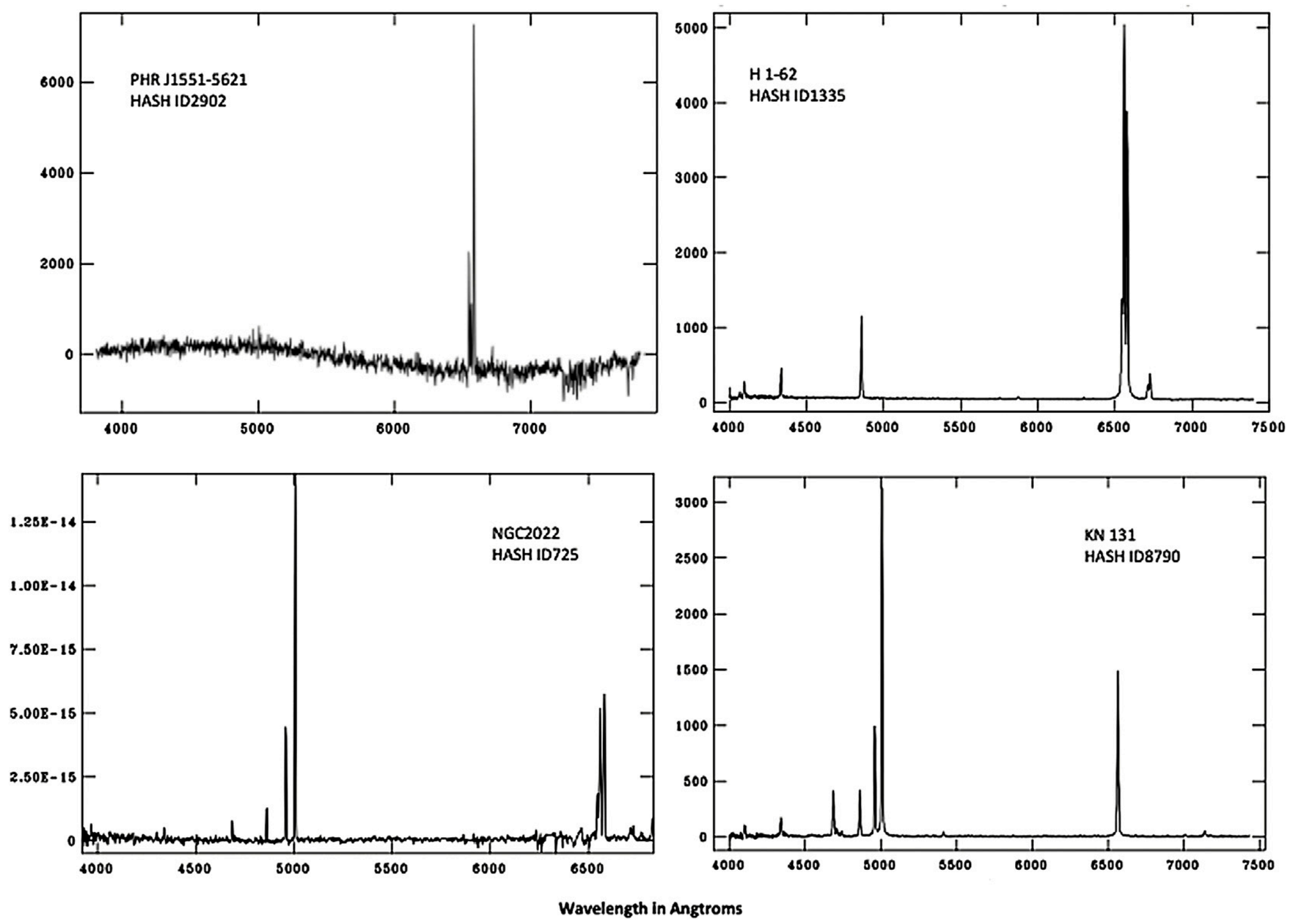

FIGURE 2
Example spectra of four bona-fide PNe from HASH that demonstrate the spectral diversity of the PNe family. The top left spectrum is of evolved bipolar PN PHR J1551-5621 (PN G325.9-01.7) that suffers from extinction in the blue, making detection of [OIII] and H$\beta$ problematic. It shows a high [NII] to H$\alpha$ ratio ( ~8) and the S/N is modest. The top right example is of a compact, VLE PN H 1–62 (PN G000.0-06.8) where there is no [OIII] detectable in the blue only the Balmer series. However, in the red the [NII] to H$\alpha$ ratio is much greater than seen in HII regions while the [SII] line ratio indicates high electron density. The bottom left is of PN KN 131 (PNG162.9-01.6), an evolved bipolar of high excitation, as evident by the prominent HeII 4686Å emission line (which is itself a strong PN diagnostic). Here the [NII] to H$\alpha$ line ratio is also > 1. This is the only spectrum here that is flux calibrated. Finally, in the bottom right, we have the high excitation elliptical shaped PN NGC 2022 (PN G196.6-10.9), where HeII is as strong as H$\beta$ and the [OIII] lines are very strong. Only H$\alpha$ is prominent in the red with very weak [NII] and [SII], a common occurrence for many high excitation PNe that also tend to be of round morphology.

Finally, apart from the impact of dust on detectable emission lines, there are many other subtler variations in PNe emission lines and their ratios (see Figure 2), that depend on the physical conditions of the object itself and the more local environment. This includes the strength of the [SII] lines relative to H$\alpha$, which is an indicator of shocked gas conditions typically seen in supernova remnants and Wolf-Rayet shells. This ratio is used as a diagnostic of shocks when it exceeds ~0.5–0.6 (Fesen et al., 1985). For an excellent tutorial on PN and H II region nebula spectroscopy see Peimbert et al. (2017).

## 5.1 PNe mimics

It is not surprising other emission objects have and can be confused with PNe from their emission line spectra alone. Emission line ratio diagnostic diagrams, as developed by Baldwin et al. (1981), can provide powerful discriminatory power to assist. These diagrams were refined for application to PNe by Frew and Parker (2010). Figure 3 provides an updated version of the log(H$\alpha$)/[NII] versus log (H$\alpha$)/[SII] plot of such an diagnostic flux ratio diagram taken from Figure 4A of Frew and Parker (2010). The established empirical limits for where most PNe fall are shown by the two main black-line tracks, as established from our updated HASH data. There is overlap between object classes so points falling in boundary regions often need additional corroborating data, including using different versions of the plot with different line ratios that can provide better object type separations, e.g., see Figures 7, 8 in Sabin et al. (2013).

Example images of some of the commonest mimics are shown in Figure 4, as taken from the HASH database. The available multi-wavelength images aid greatly in identification. See Frew and Parker (2010) for more examples, including emission line spectra of selected mimics, different examples of which are also given in Figure 5 below for convenience.

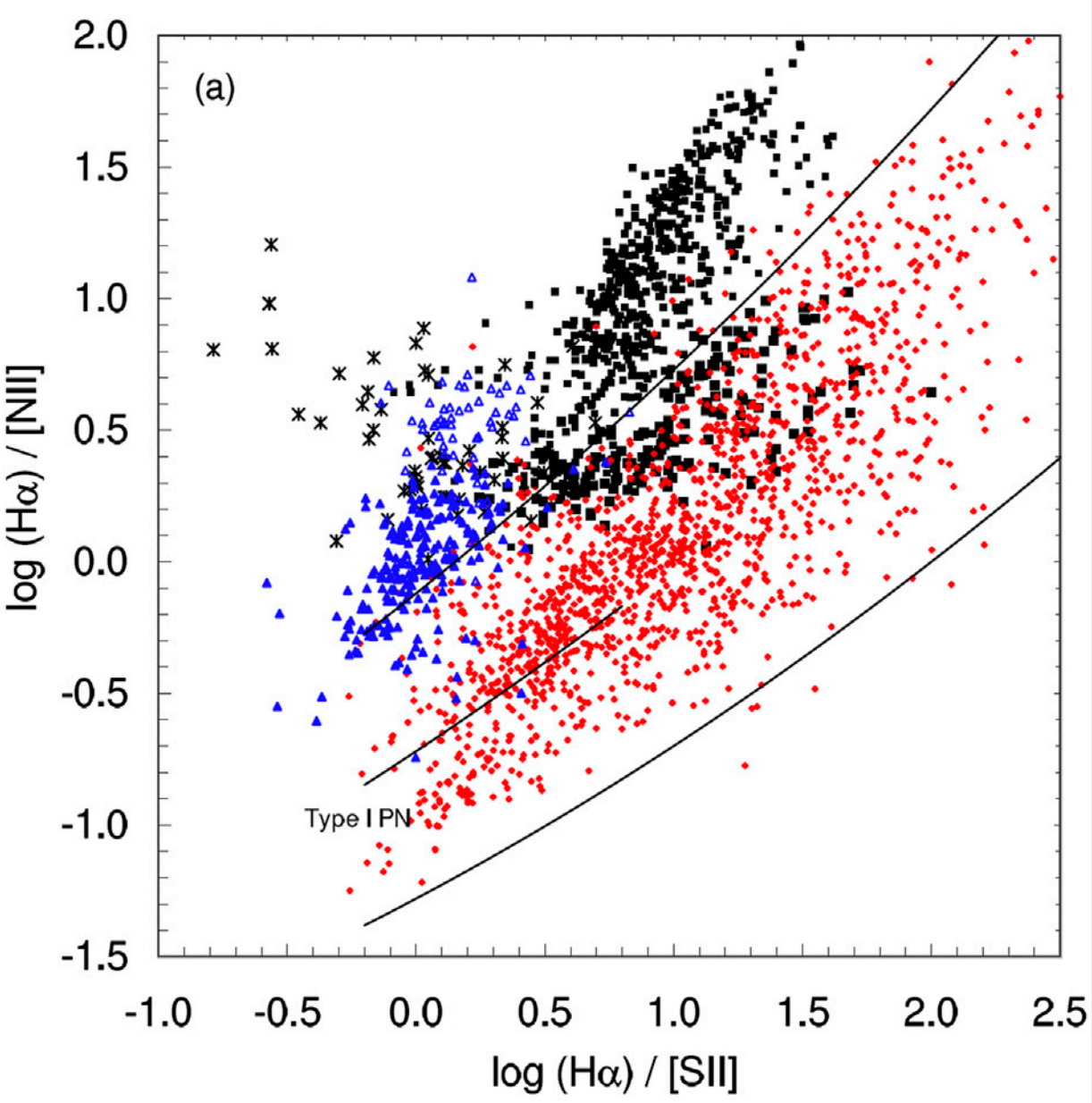

FIGURE 3
An updated version of the so-called "BPT" (Baldwin et al., 1981) diagnostic plot of log($H\alpha$)/[II] versus log ($H\alpha$)/[SII] for emission line flux ratios of various object classes with emission spectra. The plot is reproduced from panel Figure.4a of Frew and Parker (2010). Here [NII] is the flux combination of the two red nitrogen lines at 6548 and 6584Å while [SII] refers to the sum of the two red sulphur line fluxes at 6717 and 6731Å. The limits established empirically for where PN fall are shown by the two black-line tracks as established by Frew and Parker (2010) from our updated HASH data. We also plot the locii of Type I PN (as defined by Kingsburgh and Barlow (1994). There is clearly overlap between some classes so objects falling in these regions often need additional corroborating data.

In Figure 5 are shown four examples of spectra of typical PN mimics - three of which are for the same objects as the images in Figure 4. The top left spectrum is an example of ionised circumstellar matter and the top right for a more classic HII region. The spectra are very similar with $H\beta$ only in the blue - an important diagnostic. In fact the circumstellar matter example is also a HII region (Strömgren sphere) but the ionisation is by a single, isolated star and not a more general HII region which may be large, complex and host many young stars. We make this classification distinction in HASH. The strength of [NII] relative to $H\alpha$ and the [SII] lines indicate low density for the HII region. This should be compared to the PN VLE spectrum in Figure 2. The bottom left spectrum is for part of a supernova remnant (SNR). Here the strength of the [SII] 6717/67631Å lines are very strong relative to $H\alpha$, a clear indicator of shocked gas as in SNRs and Wolf-Rayet shells. Note also the presence of the 6300 and 6363Å [OI] lines typically seen in SNR but not usually PNe. The bottom right is the spectrum of a symbiotic system. Here a strong clue to nature is given by the broad $H\alpha$ line but also particularly the strength of the [OIII] 4363Å line (normally very weak in PNe) relative to $H\gamma$. It is all these different lines and their intensities and ratios, compared to the typical PNe spectra shown in Figure 2, that significantly aid in object identification when using spectroscopy.

Proper scientific exploitation of the PNe phenomena has thus been hampered by the significant numbers of non-PNe mimics that have badly contaminated previous catalogues prior to HASH. This is due to similarities in morphology and spectra between the range exhibited by PNe (refer Figures 1, 2) and those of interlopers that overlap PNe in certain characteristics (refer Figures 3, 5). These include compact HII regions, ionised ISM, Wolf-Rayet shells, Supernova remnants, Herbig-Haro nebulae, Young Stellar Objects (YSOs), reflection nebulae, low redshift emission line galaxies, nova shells and various types of emission line star (when the PNe candidate is point-like). The HASH team developed robust, multi-wavelength processes to effectively and holistically tackle this long standing problem and eliminate most interlopers (Frew and Parker, 2010)–see Section 12. This gives confidence we are working with high integrity, carefully vetted, PNe samples.

## 6 The problem of PNe distances

In any PN review the issue of determining accurate distances to Galactic PNe needs to be addressed. This is because without a distance many meaningful PN physical characteristics cannot be properly determined, including their physical size which can have a bearing on whether the nebulosity can even be a PN. This has always been an issue, apart from the rare instances when trigonometric parallaxes have been available for their CSPN, e.g., Harris et al. (2007) or through the few cases where expansion distance estimates are possible, Hajian et al. (1993). The previous lack of accurate PNe distances to more than a few dozen Galactic PNe seriously affected derivation and use of their reliable physical properties. Unfortunately, the CSPN themselves are too diverse to provide any distance information like the P-L relation for Cepheid variables (see Section 7 below). Various statistical distance techniques have been developed e.g., Daub (1982) and Cahn et al. (1992), but non are particularly reliable. That was until the advent of more dedicated studies into this issue by Stanghellini et al. (2008) and also with our own more recent, robustly calibrated PNe surface brightness radius relation, SB-r: Frew et al. (2016) that provides distances, when the appropriate trend lines are adopted, accurate to ±20%, if the accuracy of the calibrators is also assumed. Of course the Gaia astrometric satellite, e.g., Gaia Collaboration et al. (2016), now offers bench-mark PNe distances for PNe where the CSPN can be confidently identified and that fall within the $G \leq 20.7$ Gaia limits. Of course Gaia distances supersede all previous estimators in terms of accuracy. See, e.g., Kimeswenger and Barría, (2018) and Chornay and Walton (2021). Furthermore, Stanghellini et al. (2020) provide an in depth review of various statistical distance scales compared to Gaia and show that a new statistical distance scale based on Gaia calibrators is considerably better than all previous such scales. Many CSPN remain too faint for Gaia and for accurate Gaia photometry, so the best statistical distance scales above remains a valuable adjunct for PNe distance estimates for such cases.

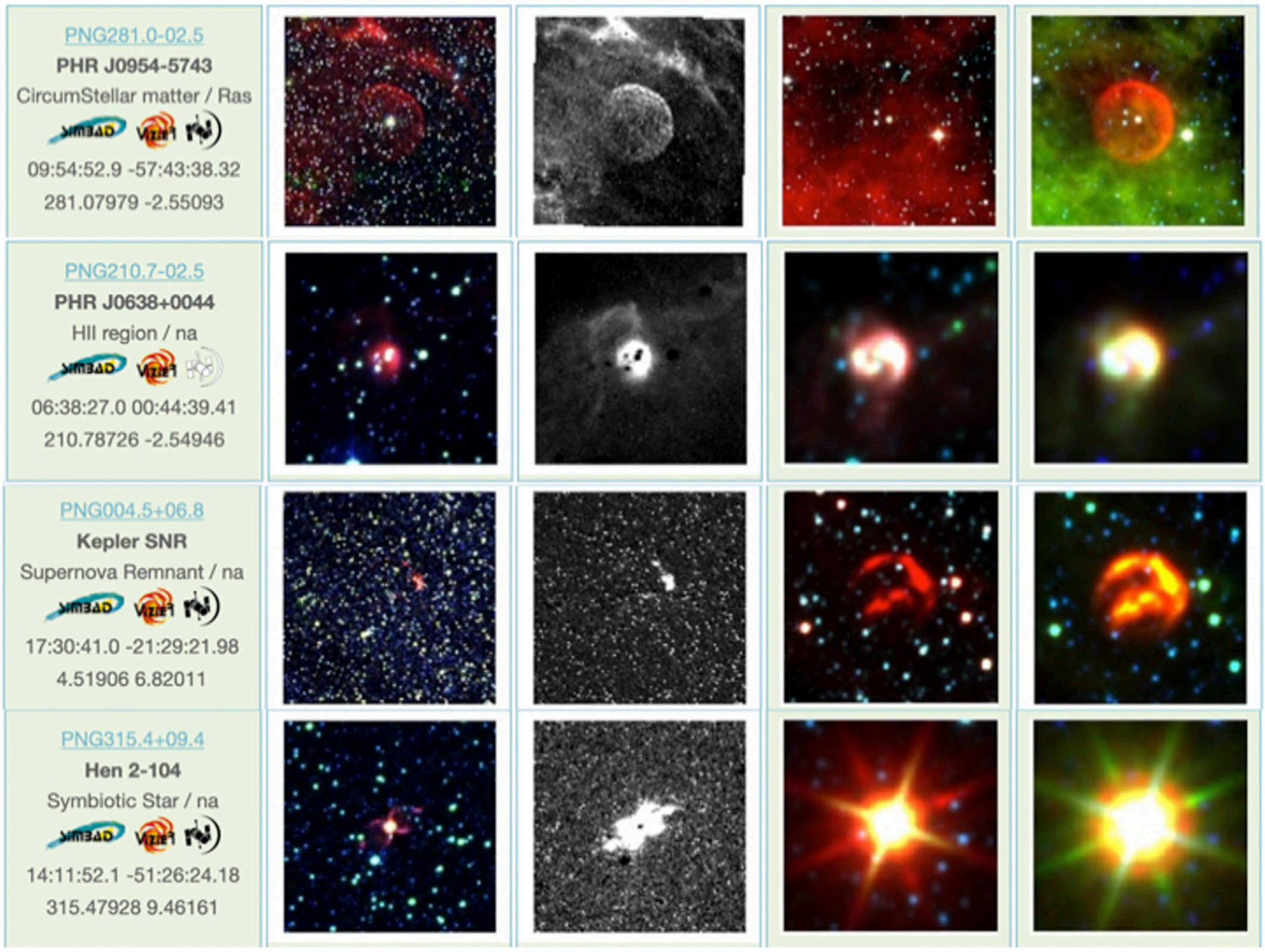

FIGURE 4
Some of the most common PN mimics. Shown from top to bottom are an example of: i) circumstellar matter; ii) a compact HII region; iii) a supernova remnant and iv) a symbiotic system. Only a small selection of the available HASH multi-wavelength imagery is shown, which is, from left to right, the SHS RGB and quotient images and the MIR WISE 321 and 432 RGB band combinations.

## 7 Central star catalogues

To help verify if a candidate nebula is a bona-fide PN the presence of a blue, hot, ionising CSPN at or near the geometric centre of the nebulosity is considered strong, positive evidence. However, until HASH, only about 20% of all known PNe had an equivocally identified CSPN (Weidmann and Gamen, 2011) to help relate PNe properties, such as morphology, ionisation state, kinematic age etc, to the underlying properties of the CSPN. As mentioned these are extremely inhomogeneous, exhibiting a wide variety of observed characteristics. These range from massive population II Wolf-Rayet stars (denoted as [WR] to distinguish them from their population I counterparts), PG1159 stars, various kinds of weak emission line stars (often denoted WELS though they are not considered to be an independent spectral classification - see Weidmann et al. (2015)), different kinds of white dwarfs (DA, DAO, DO) and early and late O(H) and Of(H) stars, e.g., Weidmann et al. (2018). A key reason CSPN are hard to locate is that they are typically of extremely low luminosity. This is because, although they are hot, they are of small physical size due to being only the residual cores of their progenitors, typically with similar diameters to our earth. When coupled with often large distances across the Galaxy this makes them hard to detect, even with large aperture telescopes. Many are beyond the photometric limits of current surveys, including Gaia.

Nevertheless, various systematic searches for CSPN have been undertaken, such as reported by Kerber et al. (2003), Weidmann and Gamen (2011), Weidmann et al. (2020) or more recently from work with the Gaia photometric catalogues, e.g., Chornay and Walton, (2021). Gaia often provides additional information on these possible CSPN, such as distances or temperatures. Preliminary work on CSPN in the HASH database was recently reported by Parker et al. (2022)

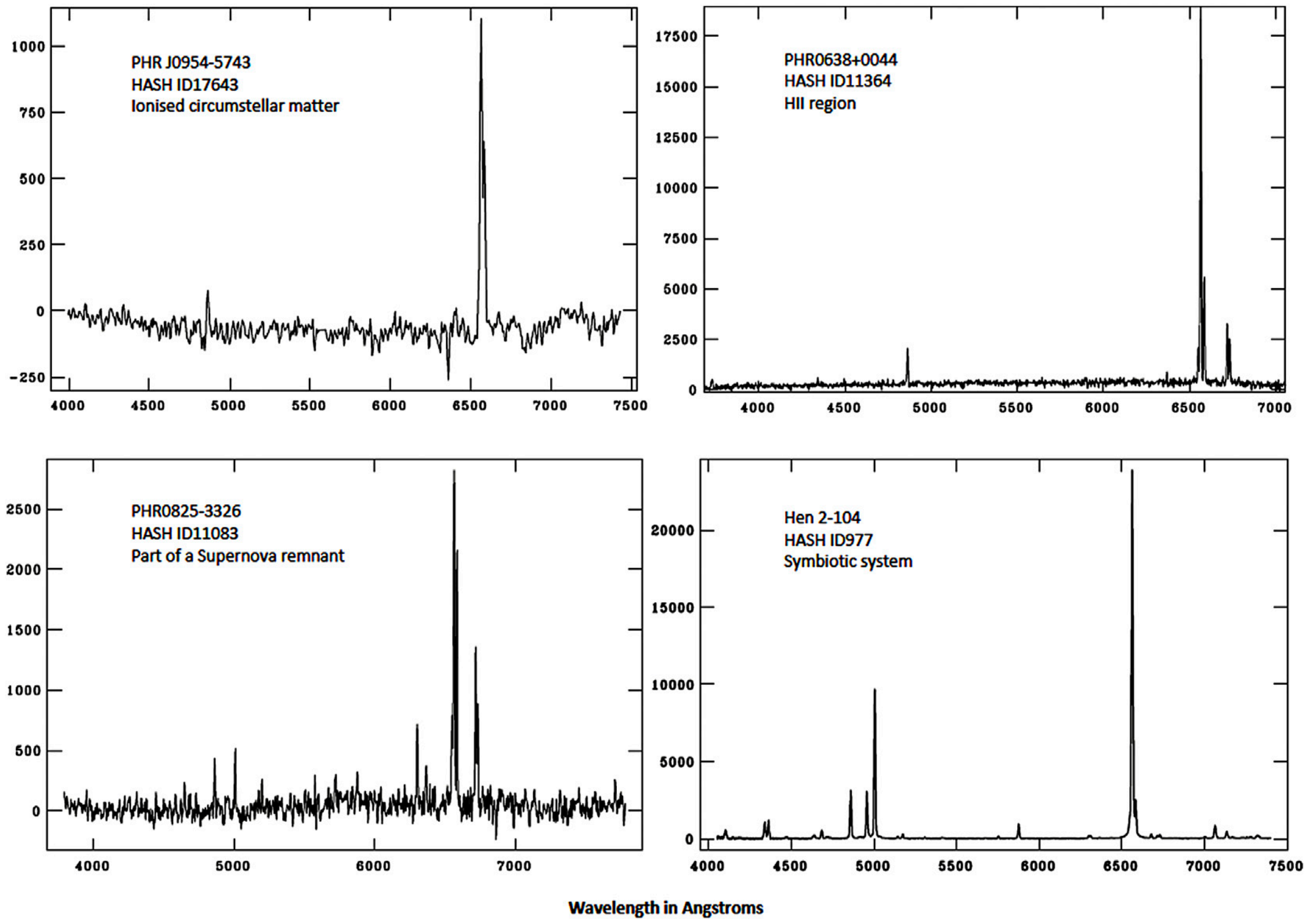

FIGURE 5
Example spectra of four mimics, three of which are the same as the images in Figure 4. Note the different lines, their intensities and ratios c.f. PNe spectra that significantly aid in object identification. The vertical axis is relative intensity.

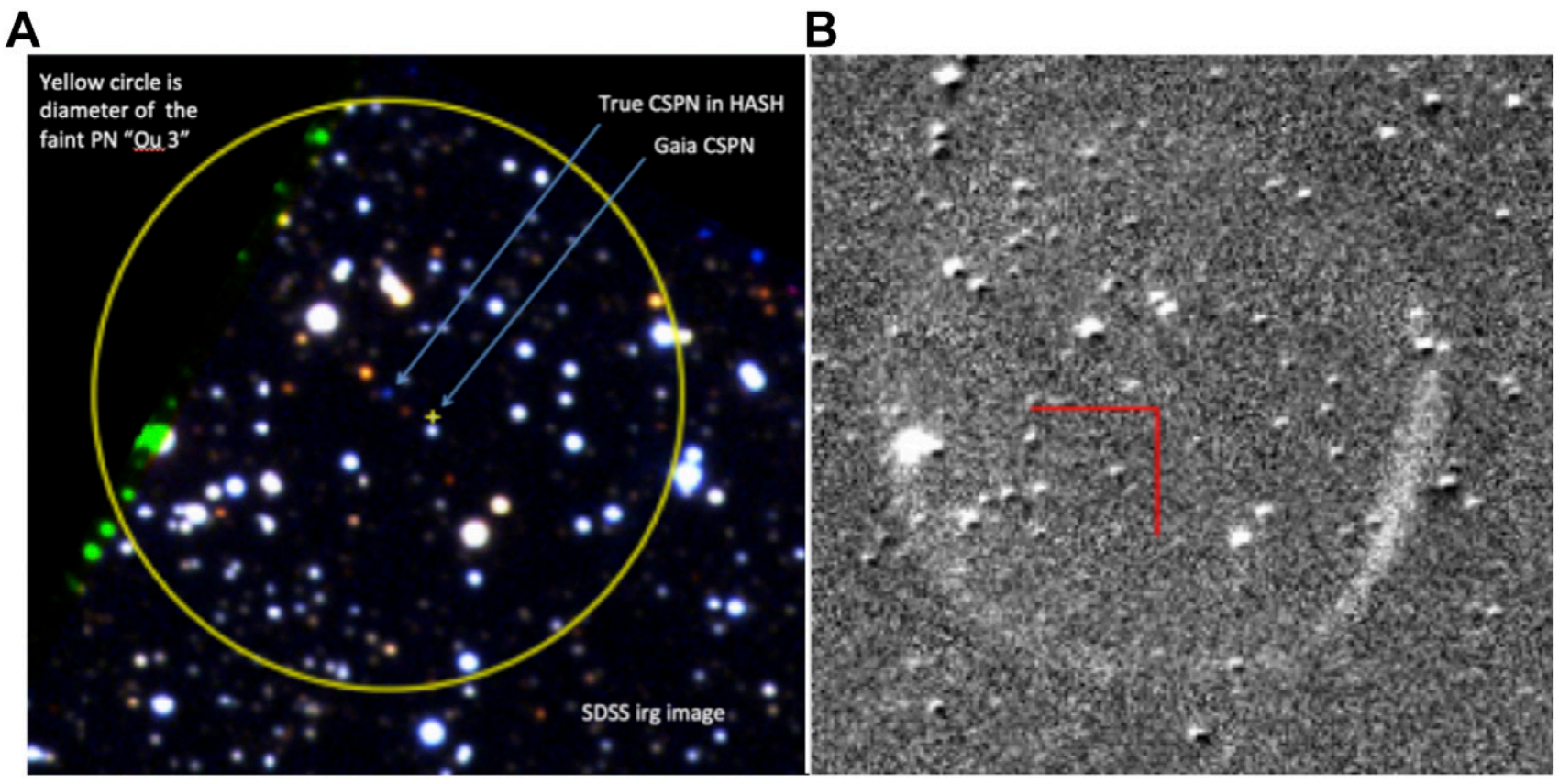

FIGURE 6
**(A)**: Annotated SDSS i,r and g band 2 × 2 arcminute RGB image of PN Ou 3 (PN G059.2 + 01.0) with North-East to top left. The incorrect CSPN identified by Gaia and the actual, faint, blue CSPN identified in HASH are arrowed. **(B)**: IPHAS Hα/R band quotient image of Ou three that shows the round 90 arcseconds diameter PN and the true CSPN location shown. This example shows the benefits of the wide variety of deep imagery available within HASH and the problems of correct CSPN identification, even with Gaia. A version of the same image has appeared in Parker et al. (2022).

which raised issues of problems of correct CSPN identification. An example of a newly discovered, faint, blue CSPN, based on imagery available in HASH, is shown in Figure 6. This example shows the benefits of the wide variety of deep imagery available within HASH and the problems of correct CSPN identification, even with Gaia. A version of the same image recently appeared in Parker et al. (2022).

## 8 The new golden age of PNe discovery

Over the last 20 years we have entered a golden age of PNe discovery. The advent of deep, wide-field, high resolution, high-sensitivity, narrow-band optical surveys centred around the prominent PNe emission line of H$\alpha$, has revolutionised our ability to trawl for Galactic PNe. Pre-eminent among these are the Galactic plane H$\alpha$ survey undertaken on the UK Schmidt Telescope in Australia (Parker et al., 2005) called the SuperCOSMOS H$\alpha$ survey (SHS) and the Isaac Newton Photometric H$\alpha$ survey (IPHAS) performed on the 2.5 m ING telescope on La Palma (Drew et al., 2005). These two H$\alpha$ surveys have provided significant Galactic and also Magellanic Cloud PNe discoveries that have more than doubled the totals accumulated by all telescopes over the previous 260 years.

Furthermore, these discoveries arise from the same base survey data from a single telescope and detector technology for each survey, offering unprecedented levels of homogeneity. Such significant numbers of PNe discoveries tend to be more evolved, of lower surface brightness, more obscured or more compact/distant than most previous discoveries. This provides a broader and more representative sampling of the true, underlying Galactic PNe population. These ~1,500 new PNe were first published in the MASH catalogues (Parker et al., 2006; Miszalski et al., 2008) from discoveries with the SHS in the southern Galactic plane (Parker et al., 2005) and from the broadly equivalent northern Galactic plane IPHAS H$\alpha$ survey (Sabin et al., 2014). Both surveys have similar ~5 Rayleigh sensitivity to H$\alpha$ emission but the SHS covers ~4000 square degrees and extends to 10° above and below the Galactic Plane, while also covering the rich Galactic bulge. The IPHAS survey, on the other hand, only covers ~1800 square degrees and goes to ±5° in Galactic latitude, though the angular resolution is better. Such an advantage is of little value for well resolved PNe. This, and the location of the rich Galactic Bulge hunting ground, is why new PNe discoveries of the southern Galactic Plane have dominated numbers.

We await significant PNe discoveries from the more recent VST Photometric H$\alpha$ survey (VPHAS+) survey (Barker et al., 2018). VPHAS+ (Drew et al., 2014) was undertaken with the 2.6 m VST telescope in Chile that also covers the Southern Galactic plane, but only to the same latitude limits of ±5° as for IPHAS in the north. This ensures that the last, great, UKST photographic survey, the SHS, still remains unique and relevant today.

## 9 The multi-wavelength revolution

As already mentioned, previous PNe compilations were highly variable in quality and integrity. This is unsurprising as they contain heterogeneous assemblages of PNe identified, misidentified and re-identified again over many decades by dozens of astronomers working with a wide variety of telescopes, detectors, resolutions, wavebands and sensitivities. Furthermore, the more recent availability of sensitive ground and space-based multi-wavelength imaging surveys of high astrometric integrity provided the basis for significant new discoveries. This new generation of wide-field, multi-wavelength surveys allow us to revisit the identity, morphologies, properties and recorded positions for most PNe in existing catalogues.

Narrow and broad band optical discovery data can now also be coupled with key ground and space-based multi-wavelength observations that provide valuable, additional diagnostic power. Multi-band optical surveys include the SSS (Hambly et al., 2001), SDSS (Gunn et al., 1998), PanSTARRS (Chambers et al., 2016) and DeGaPe (Schlafly et al., 2018); near infrared (NIR) includes: 2MASS (Skrutskie et al., 2006), UKIDSS (Lawrence et al., 2007) and VVV (McMahon et al., 2013); the mid-infrared (MIR) multi-band surveys include: IRAS (Neugebauer et al., 1984), MSX (Price et al., 2001), GLIMPSE (Spitzer/IRAC: (Benjamin et al., 2003; Churchwell et al., 2009), MIPSGAL (Carey et al., 2009) and WISE (Wright et al., 2010), while the different radio frequency surveys include the NVSS (Condon et al., 1998), MGPS2 (Murphy et al., 2007) and PMN (Griffith and Wright, 1993). Where available, GALEX (Martin et al., 2005) UV data can also be used to reveal hitherto invisible ionising stars, e.g. (Frew et al., 2011) and other properties, (Pradhan et al., 2019; Hillwig et al., 2022). At higher energies more targeted, high resolution X-ray data of PNe are being obtained with Chandra (Kastner et al., 2012; Freeman et al., 2014; Montez et al., 2015) after earlier work with ROSAT (Kreysing et al., 1992; Guerrero et al., 2000). See Guerrero (2020) for a recent review.

In Figure 7 the main HASH image page for PN Abell 21 is shown that well demonstrates the power of multi-wavelength imagery for such a large but prominent PN. See figure caption and Section 12 for further details.

Taken together all these multi-wavelength surveys have further enhanced the PN discovery potential in new ways e.g., Miszalski et al. (2011) and Fragkou et al. (2018). They have provided fresh insights into their multi-wavelength characteristics, e.g., Cohen et al. (2007; 2011), while permitting, as we have shown, the more robust elimination of contaminants that have significantly impacted the integrity of previous PNe compilations (Frew and Parker, 2010).

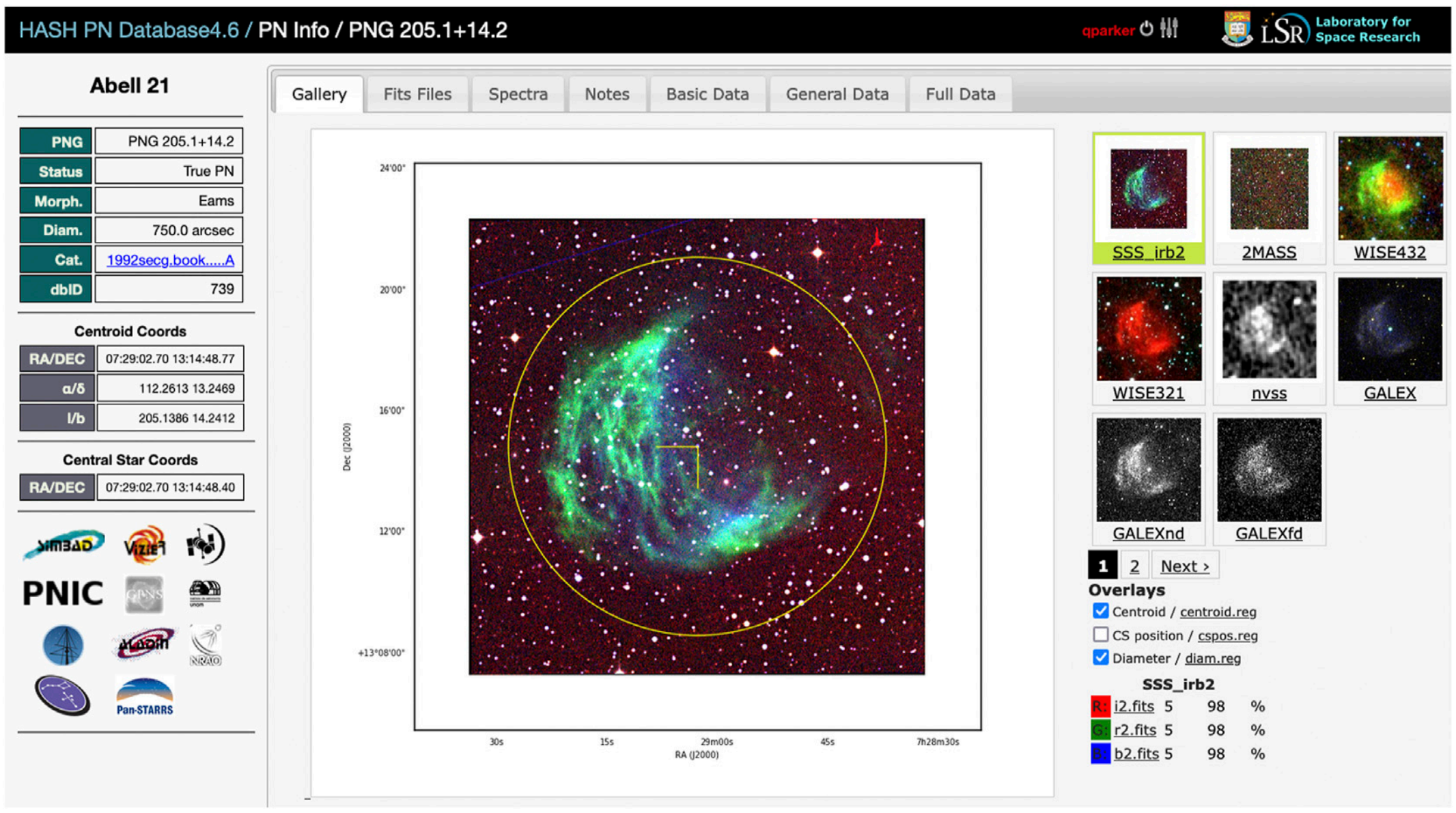

FIGURE 7
HASH image page of PN Abell 21 (HASH ID 739) that well demonstrates the power of multi-wavelength imagery for such a large but prominent PN in the broad-band optical (centre main image). To the right are the principal other multi-wavelength images available. The WISE MIR data shows an inner red region (WISE W4 band) due to the prominence of [OIV] emission at 25.89 μm, a MIR proxy for the HeII 4686Å optical emission line, showing the PN is of high excitation. The GALEX image shows the CSPN and the excitation in the surrounding gas while it is also resolved in the radio. Along the top of the HASH page for this PN are various tabs that can be selected to look at the available 1-D spectra, useful notes on the object, downloadable fits image files etc. To the upper left is the basic data including co-ordinates and lower left a clickable panel of icons that give direct access to the specific entry for the PN from SIMBAD and Vizier (data) or PanSTARRS or HST (imagery).

The advent of such high quality, non-optical surveys provides a revolution in the ability to hunt for and identify all sorts of emission line objects, not just PNe. This is especially when high levels of obscuration by intervening dust makes optical detections problematic. Such multi-wavelength data provides strong, additional discovery space and diagnostic and discriminatory power across the Galaxy. They are the basis for various PNe and other resolved emission line candidate lists in the NIR and MIR going back to IRAS for the MIR, e.g., Ramos-Larios et al. (2009) but now including Spitzer, e.g., Phillips and Ramos-Larios (2008), Mata et al. (2016) and MIPSGAL Mizuno et al. (2010) in the MIR and 2MASS for the NIR. e.g., Schmeja and Kimeswenger (2001), Ramos-Larios and Phillips, (2005) and Corradi et al. (2008) but now including VVV (McMahon et al., 2013; Weidmann et al., 2013; Minniti et al., 2019) and UKIDSS (Lawrence et al., 2007). Here specific NIR (J–H) versus (H–$K_s$) colour-colour plots provide powerful discrimination for compact PNe compared to symbiotic stars, Be stars, T-Tauri stars, Cataclysmic and Mira variables (refer Figure 3 in the Corradi et al. paper). Finally we have the radio regime, e.g., Hoare et al. (2012) and for PN candidates selected according to radio criteria Fragkou et al. (2018) and for MASH PNe, Bojičić et al. (2011). Bojičić et al. (2021) shows how new, high-resolution, high-sensitivity, multi-frequency, wide-field radio surveys such as the Australian Square Kilometre Array Pathfinder (ASKAP) Evolutionary Map of the Universe offer fresh opportunities to undertake new determinations of useful PNe parameters. This paper specifically includes an application to determine accurate angular-sizes of PNe using a new radio continuum spectral energy distribution fitting technique. This can be applied to unresolved and/or heavily obscured PNe that are extremely faint or even non-detectable in the optical so is a valuable new capability.

Using these resources we have now constructed a new type of PNe repository. It effectively federates all these data sets, catalogues, surveys and discoveries, along with their extant spectroscopy, into a single 'research platform'. This repository and database is called HASH, Parker et al. (2016) and Bojičić et al. (2017) and is described in more detail in Section 12 below. We can then investigate, re-evaluate and if necessary, re-assign, all objects currently or previously identified as PNe in our Galaxy and also hunt for their CSPN (Parker et al., 2022). Figure 8 gives an excerpt from the HASH database listing several of the current 2670 "True" Galactic PNe, in imaging mode, where user selected multi-wavelength images can be plotted for each PN, row by row.

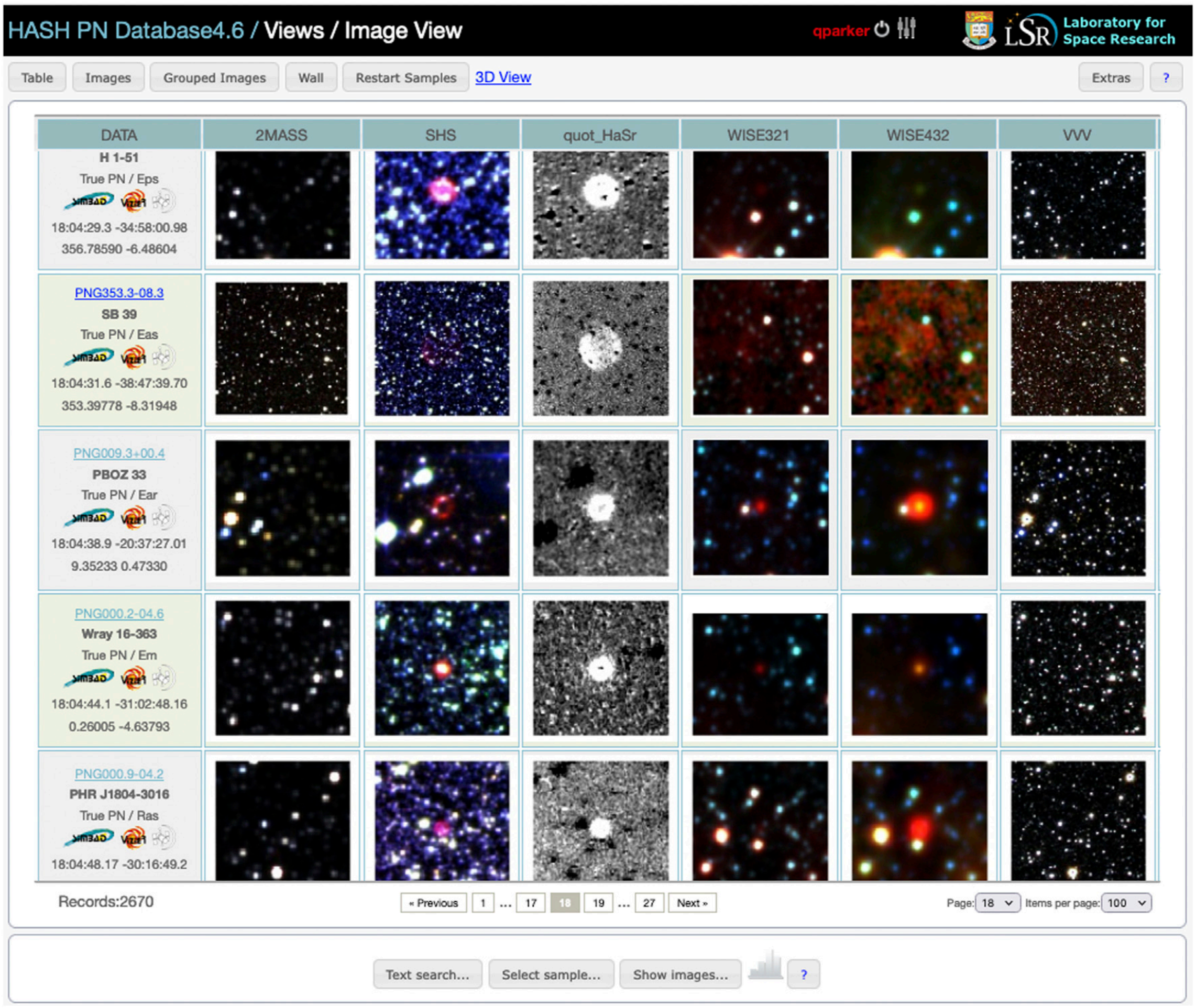

FIGURE 8
Excerpt from the HASH database of several "True" Galactic PNe in imaging mode where user selected, multi-wavelength images can be plotted for each PNe, row by row. Up to 48 such images are available for selection from 25 different surveys. Only five of the 2670 rows available here for 'T' Galactic PNe are shown.

## 10 Current PNe candidate discovery, identification and verification techniques

PNe candidate discovery processes take many forms and can include automatic photometric trawls for emission line objects from optical broad-band and narrow on/off-band survey data when the source is sufficiently compact to register. However, such apparent emission-line excess, point-like sources contain all varieties of emission line candidates and not just very compact (young) PNe. These include various types of emission line star, compact HII regions and late type stars where increasingly prominent molecular bands in the far red can fall in the narrow band and give a strong, apparent emission signal compared to the broad band. For examples see Viironen et al. (2009; 2009b). Spectroscopic confirmation is always required with such samples and success rates for finding compact PNe are low, being dominated by other types of compact emitter.

Of course most (e.g., 88%) of all "True" Galactic PNe in HASH are well resolved and can be of complex morphological structure with variable surface brightness (refer Table.1 in Section 12). They can be most easily detected in narrow-band imaging surveys of various kinds in the optical (usually [OIII] and H$\alpha$) and in the MIR and radio wavelengths in particular (Parker et al., 2012). PNe identification however, remains a complicated process, notwithstanding the various problems with mimics described earlier. This is because of the wide variety of morphologies, ionization characteristics and surface brightness distribution exhibited by the broad PNe class and their main location in the often extremely crowded and obscured Galactic Plane. These variables reflect the stage of nebular evolution, progenitor mass and chemistry and the various shaping influences previously mentioned (common envelope binaries, magnetic fields, ISM interactions, sub-solar planets). Then there are the effects of interstellar extinction and the broader interstellar environment where source confusion in dense star fields like the Galactic Bulge can be problematic.

Nevertheless, PNe candidates continue to be discovered on a regular basis and in surprisingly significant numbers. This is increasingly by the dedicated amateur community (see Section 11 below) thanks to the availability and proliferation of on-line digital multi-wavelength surveys and via the subsequent different discovery pathways now available, and, of course, serendipitously. In some cases the same mistakes and biases that have bedeviled the pre-MASH and HASH catalogues remain an issue. The discoverer is encouraged to follow the detailed evaluation work of Frew and Parker (2010) and also to check positions against entries in HASH to verify whether a "discovery" is in fact new.

## 11 The value of the amateur community in finding PNe

One increasingly important avenue adding significantly to the Galactic PNe population is now coming from the amateur community, especially over the last 10 years. They have adopted several key PNe discovery techniques. The primary one is to trawl systematically the "out of plane" on-line, broadband optical surveys looking for unregistered, low-surface brightness nebulosities at larger Galactic scale heights. The second technique is to go over existing on-line, narrow band Galactic plane surveys such as the SHS and IPHAS to hunt for candidates missed by the professional MASH/IPHAS teams. A third process has been to search the WISE survey for resolved MIR nebulae that could be PNe. The original but now largely inactive Deep Sky Hunters (DSH) team, led my Matthias Kronberger, is an excellent example of what can be achieved by trawling such datasets (Kronberger et al., 2012, 2014, 2016). Candidates are followed up with deep narrow-band images on a series of amateur and professional telescopes.

Another more prominent and now very active group is led by Pascal Le Dû, in France, e.g. Acker and Le Dû, 2014 and Le Dû et al. (2018) who employ similar techniques to great effect. The main project description paper is LeDû et al., 2022, A&A (in-press). They also perform their own spectroscopic confirmation of many new candidates on a range of bespoke spectroscopic facilities on a suite of amateur telescopes of reasonable aperture. This group has developed a comprehensive web site that details these discoveries[1]. As at March 2022 their efforts have uncovered 210 spectroscopically confirmed PNe with 123 True (T), 51 Likely (L) and 36 Possible (P) Galactic PNe, all ingested into HASH and representing ~5% of all known Galactic PNe. A further 610 PNe candidates await follow-up.

Finally, the amateur community is also making use of dedicated small aperture and sometimes automated telescopes on sites in Chile and Australia. This is to perform very deep and lucky narrow-band imaging of many PNe, taken, in some cases, over dozens of hours. These observations are used to provide unprecedented, high-quality, deep imaging, rivaling and in many cases surpassing, the best professional images of these PNe. An excellent example of what is available can be found on the Astrodon site[2]. Two selected examples are shown in Figure 9 that demonstrates the power and value of this amateur work.

## 12 The "HASH" consolidated PNe database and research platform

All SHS (Parker et al., 2006; Miszalski et al., 2008) and IPHAS PNe discoveries (Sabin et al., 2012), together with all previously published PNe and other new, smaller samples independently published by other groups, including those of the active amateur community, have been compiled and incorporated into the so-called "HASH" catalogue and research platform (Parker et al., 2016). HASH has so far ingested 109 published PN catalogues, compilations and lists with new discoveries added as they are made known. All such discoveries and previously published PNe were independently re-measured and vetted based on the overall body of data and evidence (see Section 12.1 below).

The HASH database[3] is currently maintained by the Laboratory for Space Research (LSR) at the University of Hong Kong and is considered the final arbiter of PN identification. It contains 11,460 entries as of March 2022 including over 3,830 True (T), Likely (L) and Possible (P) PNe in the Galaxy and 821 T, L, P PNe in the Magellanic Clouds. HASH includes ~500,000 fits cutouts and ~88,000 colour images from up to 48 different image types from 24 major multi-wavelength surveys. In Table 1 the current major morphological breakdown for the PNe HASH content is presented.

The remaining entries consist of: i) the various mimics that have at one time or another been classified as PNe, including plate artifacts and image flaws; ii) other emission nebulae and compact objects found by the MASH and IPHAS teams but not considered PNe; iii) other object types spread across 41 different classifications and finally: iv) 2400 + candidates still requiring follow-up and classification. Some of the most interesting and numerous mimics are listed in Table 2, while examples of their typical emission line spectra are given in Figure 5 and in Figure 3 of Frew and Parker, (2010).

1 http://planetarynebulae.net/FR/

2 https://astrodonimaging.com/gallery_category/nebula/

3 http://hashpn.space/

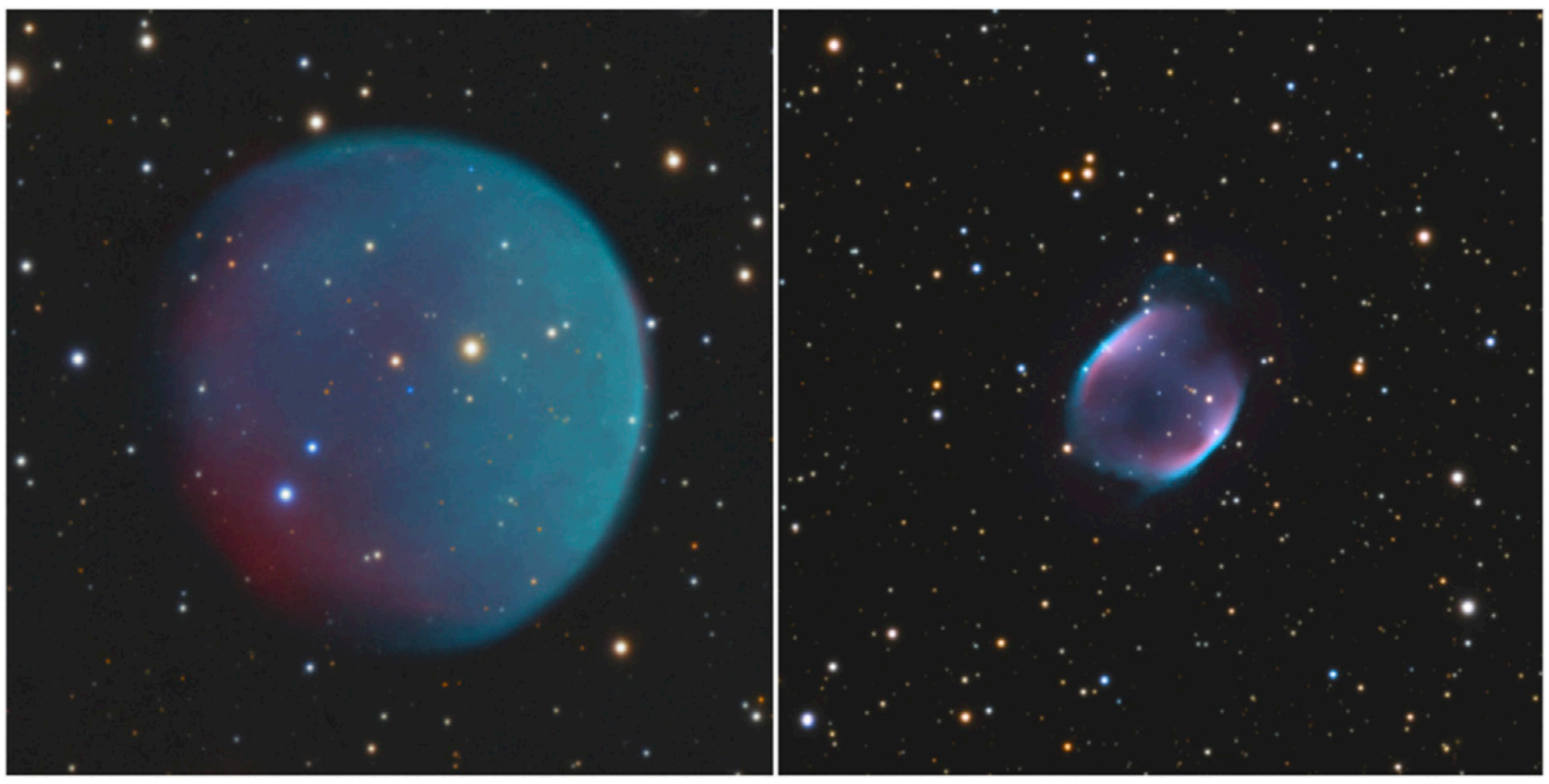

FIGURE 9
Deep amateur multi-narrow band RGB images of two recently confirmed amateur PN discoveries StrDr 140 (HASH ID 33480) and StDr 141 (HASH ID 33483). A faint blue CSPN is evident in both. Left: StDr 140 with a diameter of 414 arcseconds. Imaging by Peter Goodhew with processing by Marcel Drechsler. Total exposure time: 76 h; Hα 150 × 900 s; [OIII]: 125 × 900 s. Right: StDr 141 with a main axis size of 252 arcseconds. Imaging by Peter Goodhew with the CSPN at geometric centre. The PN is a bipolar with a strong oval core. It is much stronger in [OIII] than Hα.

TABLE 1 Breakdown of all HASH T (true), L (likely) and P (possible) PNe by base morphological type (shape) and the number of spectra available for each status. These base types are E: Elliptical/oval, R: Round, B: Bipolar; I: Irregular; A: Asymmetric; S: Star-like/compact or apparent point source.

| PN status | Shape | | | | | | Unassigned morphology | Spectra available |
|---|---|---|---|---|---|---|---|---|
| | E | R | B | I | A | S | | |
| T 2670 | 1058 | 563 | 668 | 20 | 10 | 329 | 22 | 2475 |
| L 464 | 174 | 70 | 39 | 11 | 12 | 81 | 67 | 260 |
| P 696 | 152 | 43 | 28 | 14 | 21 | 126 | 312 | 205 |

TABLE 2 List and numbers of the most common PN mimics from among the 41 different non PNe classifications used in HASH.

| PN mimic | Number found | PN mimic | Number found |
|---|---|---|---|
| HII region | 304 | Star | 249 |
| Star Cluster | 235 | Symbiotic stars | 216 |
| Galaxies | 203 | Circumstellar matter | 167 |
| Artifacts/plate flaws | 164 | Emission line star | 142 |
| SNR/SNR candidates | 142 | PAGB/Pre-PNe | 125 |
| RV Tau | 111 | Reflection Nebulae | 98 |
| YSO/YSO candidates | 60 | Ionised ISM | 51 |

This enormous undertaking by the MASH and now HASH team over 22 years is on-going. HASH will also soon include accurate [OIII] (Kovacevic et al., 2011) and Hα fluxes (Frew et al., 2013, 2014) for large numbers of PNe and where available distances either from Gaia or via our surface-brightness radius relation, Frew et al. (2016), that was used to construct Figure 1. Gaia (Gaia Collaboration et al., 2016) will help for those PNe where the CSPN has been identified and is brighter then the Gaia magnitude limits ($G \leq 20.7$). The HASH team has also catalogued thousands of interesting non-PNe and re-assigned significant numbers of supposed PNe into other object-types based on our robust identification techniques and better data (Frew and Parker, 2010). All these activities, outputs and experiences have been combined and incorporated into HASH in one form or another. Figure 10 shows the percentage distribution of all T, L, P Galactic HASH PNe in pie-chart form, as well as where all the HASH PNe have broadly originated. Here 'ESO' refers to the Acker PNe catalogues, 'Kouhoutek' to the various Perek and Kohoutek compilations with the other designations self-evident.

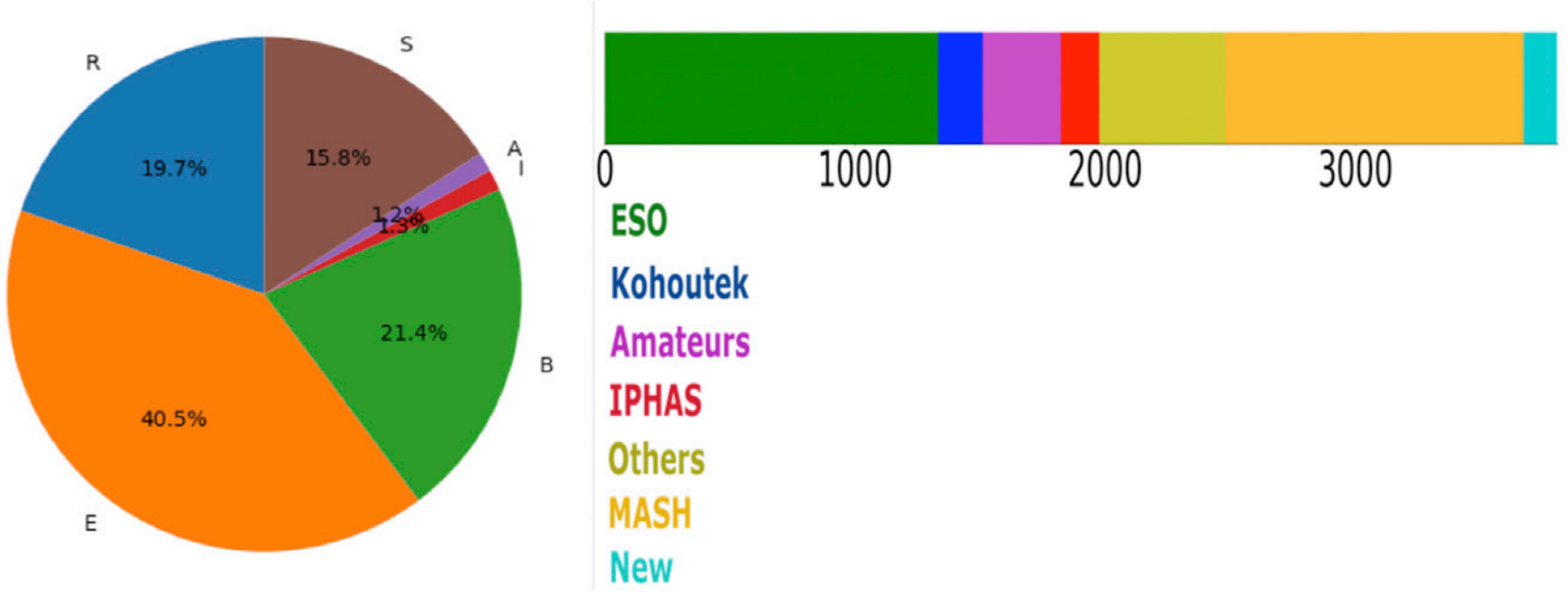

FIGURE 10
Graphical representation of the morphological percentages of Galactic PNe in HASH and where the HASH PNe have originated. On the right hand side bar chart of PNe listing origin, "ESO" refers to the Acker et al. PNe catalogues, "Kouhoutek" to the various Perek and Kohoutek compilations while the other designations are self-evident.

Consequently, HASH provides, for the first time, an accessible, reliable, on-line, SQL database for essential, up-to date information for all known Galactic and even Magellanic Cloud PNe. It is effectively a one-stop research platform for studying PNe. In constructing HASH we have attempted to: i) reliably remove PN mimics/false ID's that have biased previous studies; ii) provide accurate key parameters (position, angular size, morphology etc), iii) make available multi-wavelength imagery and spectroscopy wherever possible. Links to CDS/Vizier for the archival history of each object and other valuable links to external data repositories and surveys are also provided. These links include SIMBAD[4], Vizier[5], Aladin[6], all from the CDS; PanSTARRS, HST imagery, MAST[7], PNIC[8], NRAO and the SPM kinematic database (Richer et al., 2010). With the HASH interface, users can sift, select, browse, collate, investigate, download and visualise the entire currently known Galactic and Magellanic Cloud PNe inventory in all their diversity. HASH provides the community with the most complete and reliable data with which to undertake new science and is strongly recommended for use by any worker in the field. Please see our website "hashpn.space" and register for an account to use HASH. Further details of HASH are given by Parker et al. (2016) and Bojičić et al. (2017).

4 https://simbad.u-strasbg.fr/simbad/
5 https://vizier.cds.unistra.fr/viz-bin/VizieR
6 https://aladin.cds.unistra.fr/aladin.gml
7 https://archive.stsci.edu/dss/
8 https://faculty.washington.edu/balick/PNIC/

## 12.1 Confirming an object as a PN−the HASH holistic approach

When making a final decision on the nature of any PN candidate for inclusion in HASH, we attempt to apply a consistent and holistic approach based on assessment of all the available evidence and data (see Section 12.2 below). This includes multi-wavelength imagery, spectral characteristics, multi-wavelength photometry, presence of a plausible CSPN, environmental and morphological considerations, object angular/physical size and other factors. Whether in the light of all these available data an object is considered as a T, L or P PNe or indeed as some other object type, depends on how much of this evidence falls in favour of being compatible with a PN identification and at what level of confidence. This assessment includes considering all the pros and cons of the various data indicators where some of these diagnostics have higher weight than others. This entire process has been carefully established in the comprehensive discussion in Frew and Parker (2010) which I do not propose to repeat here. However, a newly developed summary of the key data that forms the decision tree in Figure 11 for assigning object status is provided below for convenience and completeness for this review.

## 12.2 Concise list of typical data sets used to help confirm PN identification for HASH

An ideal wish list of the key data needed for confirming any nebulous object as a bona-fide Galactic PN would comprise the following 10 key factors, listed in perceived order of importance and usefulness. The more these points are satisfied the more confidence in the PN identification. There are a few more minor factors (see below) but these are the most important. For all these it is assumed high quality data are available. When this is not the case the reliability of the indicator is reduced. The process is not always equivocal and can be more subjective as the quality and available indicators declines. The top four are the most simple and influential traditionally and, if available at the right quality, allow robust confirmation as a "True" PN when now combined with item 5. If some of the top indicators are absent or of lower quality or indeed ambiguous, then the candidate is assigned as a "Likely" PN. If only limited supportive evidence is available, for example due to poor data or conflicting or ambiguous indicators, then a 'Possible' designation is assigned. Finally, if there is clear evidence that the candidate is another class of object, even if some of the other characteristics are PNe compatible, than an alternative object type identification is assigned.

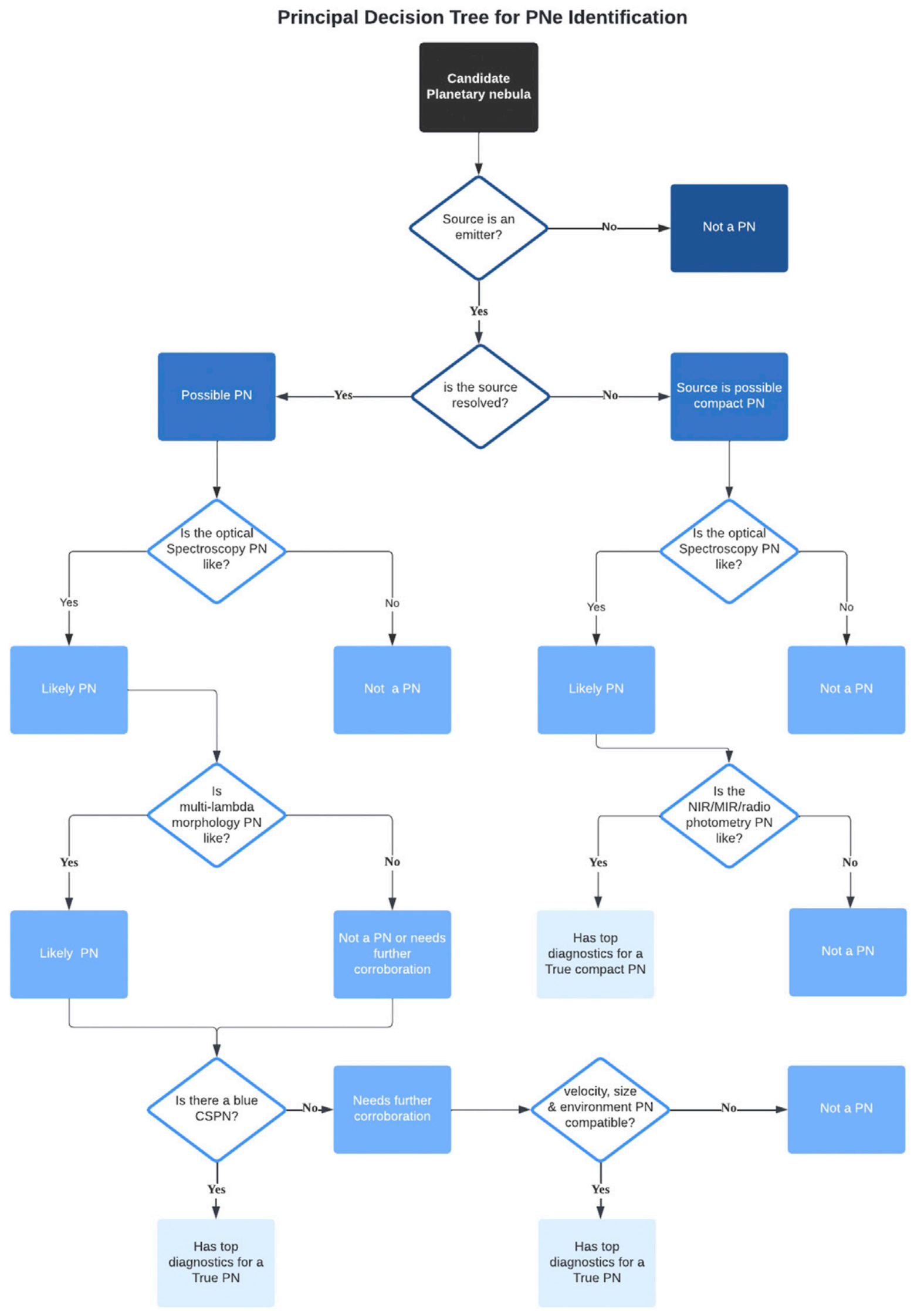

FIGURE 11
Principal HASH PN identification decision tree in the form of a basic flow-chart split into two main branches depending on whether the candidate is resolved or compact in available imagery. This chart satisfies most eventualities when good data is available for all key diagnostics. For cases where the data is not clear cut and/or evidence is contradictory, further corroboration is needed and candidates are then classed as likely or possible PNe if there are no clear grounds for complete rejection (see main text).

1. Narrow-band imaging, revealing the object is an emitter compared to broad-band equivalents. This removes confusion with reflection nebulosity and is a pre-requisite for any subsequent PN identification.
2. Clear, canonical or indicative optical morphology (can be from narrow or broad band), classified according to our ERBIAS/sparm scheme adopted for HASH (Parker et al., 2006) and discerned from good quality imagery. A few PNe are shaped by ISM interactions Sabin et al. (2012) but this is rare in all extant samples. Many mimics can be removed at this level.
3. Decent S/N spectroscopy with emission lines compatible with a PN. Such lines can vary at any given evolutionary stage and type. e.g., lack of [OIII] in VLE PNe, presence of HeII 4686Å in high excitation PN (a powerful PN diagnostic), strong [NII] to Hα in Type I PNe etc (refer Figure 2). All lines should respect the usual intensity range and ratios for common PNe lines and usual diagnostic indicators (refer Figure 3). These can reveal mimics, including Wolf-Rayet shells and Supernova remnants (where the [SII] to Hα ratio can be large), various kinds of emission line stars if the candidate is point-like, and Symbiotic systems where the [OIII] 4363Å line can rival Hγ in intensity. This line is weak in PNe.

4. Presence of a plausible ionising star (CSPN) at or near the geometric centre of the nebula. It is usually assumed this star is the true CSPN at the same distance as the PN. It should be of the right type (refer Section 7), luminosity etc to explain the ionisation in the nebula spectrum. Superposition of unrelated stars in the crowded Galactic plane is a problem (Parker et al., 2022). GALEX UV data can provide additional corroboration for a hot CSPN while GAIA can provide distances if the CSPN falls within the GAIA limits and can be correctly identified.
5. Multi-wavelength imagery and photometry that supports the optical imagery. The HII region nature of many mimics becomes evident just from the MIR imagery alone (e.g. see Figure 4, 2nd row). For some highly obscured PN candidates MIR or high resolution radio imagery can also reveal supportive morphological detail. For compact PNe candidates NIR (Corradi et al., 2008) and MIR colour-colour selections are particularly useful:
   a. In the NIR where for compact candidates genuine PNe are located in a well defined region in an I, J, $K_s$ diagram, approximately in the range $0.0 \le (I–J)_0 \le 1.5$ and $0.5 \le (J–K)_0 \le 1.5$, well separated from symbiotic Miras, see Figure 3. in Schmeja and Kimeswenger (2001) and Corradi et al. (2008).
   b. In the MIR via photometric colour selections for PNe which seem robust, see Parker et al. (2012) and Figure 3 in that paper. These are based on selections in the [3.6]−[4.5] versus [5.8]−[8.0] micron IRAC colour-colour plane. PNe fall in a well defined colour range on each axis of ~0.6 to 1.1 for [3.6]−[4.5] and ~1.65 to 2.1 for [5.8]−[8.0].
6. MIR/radio properties that also fall within the PN domain e.g., Cohen et al. (2011) where the median 8 micron MIR to radio (NVSS/MGPS2) flux ratio for all PNe is 4.7 ± 1.1. This does not vary much with PN evolutionary phase. Such photometry includes any detectable putative CSPN or symbiotic system (see Figure 4, 4th row) that are screened.
7. Kinematic properties compatible with known PNe. Typical expansion velocities are only 25-30 km/s with only a small dispersion and few exceptions–e.g., Robinson et al.. (1982) and Gesicki and Zijlstra (2000).
8. Radial velocities from emission lines indicating the object is actually Galactic (so ±300–400 Km/s) and not a nearby, compact emission line galaxy where modestly redshifted Hα lines ( < 1,200 km/s) can still fall within the narrow-band filter bandpass of Hα surveys.
9. Physical size(s) that are plausible for PN even when highly evolved. Sizes of 3 pc are now known (Pierce et al., 2004) but not larger. This depends on knowing the distance to the PN.
10. The object is found in a suitable Galactic environment for hosting PNe, i.e., not a young, dusty, HII region prone zone where compact HII regions are common mimics. If a good candidate is found then it could be foreground to that environment. Here strong data support from at least the first three points above must be present.

In reality, one or more of these desirable data sets are usually missing or of insufficient clarity or quality or even inapplicable (e.g., no point source photometry for resolved PNe except, perhaps, for their CSPN if bright enough). Decisions are then made by looking closely at all the available evidence. This inevitably remains a somewhat subjective process, especially when the data quality is mixed and some of useful indicators not clear cut. To assist in this process all the important elements of the HASH decision tree are provided in Figure 11. It has two main branches depending on whether the PN candidate is resolved or compact in the currently available imagery. This chart satisfies the great majority of eventualities well, particularly when good data is available for all key diagnostics. For cases where the data is not clear cut and/or evidence is contradictory, further corroboration is needed. This can take the form of deeper narrow-band imagery in key emission lines, deeper spectroscopy to see more diagnostic lines, identification and determination of spectroscopic and photometric properties of any CSPN that support PNe diagnosis, high resolution radio mapping etc. In the meantime candidates are classed as likely or possible if there are no clear grounds for complete rejection and re-assignment to another object type or assigned as a mimic depending on the preponderance of positive indicators. All our classifications are on a best efforts basis from the currently available data and decades of accumulated experience. As new data becomes available objects continue to get re-assessed and re-assigned as appropriate. Hence, HASH remains a living, dynamic platform at present.

## 13 The future of PNe catalogues and discoveries

As of March 2022 HASH has ~460 users from more than 60 countries and with 250 + affiliations. For the moment, the HASH database remains the most valuable 'Virtual Observatory' compliant repository for Galactic and Magellanic Cloud PNe for facilitating PNe based research. The new PNe database and interface provided by the French amateur community for all their confirmed and several hundred remaining PNe candidates is also very much worth consulting too[9]. We are now well into the

9 http://planetarynebulae.net/FR/

multi-wavelength, large-scale survey era, exemplified by the great ESO public surveys on the VISTA and VST telescopes such as VPHAS+, VVV and on-going surveys with Skymapper in Australia and those planned, including with the LSST, where temporal variability becomes a key, new element of diagnostic power. The currently available, tried and tested PNe candidate discovery and evaluation techniques, described herein, are based on access to such surveys and will continue and be further honed. These will further provide, in combination with new imaging and photometric datasets currently envisaged, many additional discoveries. An even more complete Galactic PNe inventory in the years to come is promised.

## Author contributions

The author QAP is solely responsible for writing this paper and agrees to be accountable for the content of the work.

## Funding

QAP thanks the Hong Kong Research Grants Council for GRF research support under grants 17326116 and 17300417.

## Conflict of interest

The author declares that the research was conducted in the absence of any commercial or financial relationships that could be construed as a potential conflict of interest.

## Publisher's note

All claims expressed in this article are solely those of the authors and do not necessarily represent those of their affiliated organizations, or those of the publisher, the editors and the reviewers. Any product that may be evaluated in this article, or claim that may be made by its manufacturer, is not guaranteed or endorsed by the publisher.

## Acknowledgments

QAP is grateful for the support and effort of many students, postdoctoral fellows and colleagues over the last 20+ years in their various contributions to the MASH and then HASH catalogues of Galactic Planetary Nebulae. The most generous souls amongst a large group are, in approximate chronological order of involvement: Malcolm Hartley, Delphine Russeil, Agnes Acker, James Marcout, Alan Peyaud, Mark Pierce, Rhys Morris, Ella Hopewell, David Frew, Ivan Bojičić, Warren Reid, Jayne Birkby, Brent Mizalski, George Jacoby, Albert Zijlstra, Anna Kovacevic, Kyle De Pew, Ashkbiz Danekhar, Rozenn Boissay, Lizette Ramirez-Guzmann, Danica Draskovic, Travis Stenborg, Andreas Ritter, Laurence Sabin, Vasiliki Fragkou, Claire Lykou, Kris Akira Stern, Kamila Chan, and Shuyu Tu. The author also thanks the following key members of the amateur community for our pro-am collaboration that is proving so fruitful: Pascal Le Dû, Matthias Kronberger, Marcel Drechsler, Sakib Raskol, Dana Patchick, and Peter Goodhew and for use of the deep amateur images of StDr 140 and StDr 141. Particular thanks must go to my IT research staff and postdoctoral fellows Ivan Bojičić and Andreas Ritter for their huge efforts in putting the HASH database together at HKU under my guidance. I also wish to express a special debt of gratitude to David Frew who accompanied me on much of the earlier journey as evidenced by the many references to his/our work herein and who has himself made very significant contributions to the field. I apologise if I have inadvertently omitted any deserving names from this list. Finally, I would like to thank the two referees who have helped improve this manuscript.